\documentstyle[twoside,epsf]{article}

\catcode`\@=11
\long\def\@makefntext#1{
\protect\noindent \hbox to 3.2pt {\hskip-.9pt
$^{{\eightrm\@thefnmark}}$\hfil}#1\hfill}       

\def\@makefnmark{\hbox to 0pt{$^{\@thefnmark}$\hss}}    

\def\ps@myheadings{\let\@mkboth\@gobbletwo
\def\@oddhead{\hbox{}
\rightmark\hfil\eightrm\thepage}
\def\@oddfoot{}\def\@evenhead{\eightrm\thepage\hfil
\leftmark\hbox{}}\def\@evenfoot{}
\def\sectionmark##1{}\def\subsectionmark##1{}}

\newcounter{sectionc}\newcounter{subsectionc}\newcounter{subsubsectionc}
\renewcommand{\section}[1] {\vspace{12pt}\addtocounter{sectionc}{1}
\setcounter{subsectionc}{0}\setcounter{subsubsectionc}{0}\noindent
    \par\vspace{5pt}}
\renewcommand{\subsection}[1] {\vspace{12pt}\addtocounter{subsectionc}{1}
    \setcounter{subsubsectionc}{0}\noindent
    {\bf\thesectionc.\thesubsectionc. {\kern1pt \bfit #1}}\par\vspace{5pt}}
\renewcommand{\subsubsection}[1] {\vspace{12pt}\addtocounter{subsubsectionc}{1}
    \noindent{\tenrm\thesectionc.\thesubsectionc.\thesubsubsectionc.
    {\kern1pt \tenit #1}}\par\vspace{5pt}}

\newcounter{appendixc}
\newcounter{subappendixc}[appendixc]
\newcounter{subsubappendixc}[subappendixc]
\renewcommand{\thesubappendixc}{\Alph{appendixc}.\arabic{subappendixc}}
\renewcommand{\thesubsubappendixc}
        {\Alph{appendixc}.\arabic{subappendixc}.\arabic{subsubappendixc}}

\renewcommand{\appendix}[1] {\vspace{12pt}
        \refstepcounter{appendixc}
        \setcounter{figure}{0}
        \setcounter{table}{0}
        \setcounter{lemma}{0}
        \setcounter{theorem}{0}
        \setcounter{corollary}{0}
        \setcounter{definition}{0}
        \setcounter{equation}{0}
        \renewcommand{\thefigure}{\Alph{appendixc}.\arabic{figure}}
        \renewcommand{\thetable}{\Alph{appendixc}.\arabic{table}}
        \renewcommand{\theappendixc}{\Alph{appendixc}}
        \renewcommand{\thelemma}{\Alph{appendixc}.\arabic{lemma}}
        \renewcommand{\thetheorem}{\Alph{appendixc}.\arabic{theorem}}
        \renewcommand{\thedefinition}{\Alph{appendixc}.\arabic{definition}}
        \renewcommand{\thecorollary}{\Alph{appendixc}.\arabic{corollary}}
        \renewcommand{\theequation}{\Alph{appendixc}.\arabic{equation}}
   \noindent{\tenbf Appendix \theappendixc. #1}\par\vspace{5pt}}
\newcommand{\subappendix}[1] {\vspace{12pt}
        \refstepcounter{subappendixc}
        \noindent{\bf Appendix \thesubappendixc. {\kern1pt \bfit #1}}
        \par\vspace{5pt}}
\newcommand{\subsubappendix}[1] {\vspace{12pt}
        \refstepcounter{subsubappendixc}
        \noindent{\rm Appendix \thesubsubappendixc. {\kern1pt \tenit #1}}
        \par\vspace{5pt}}

\newcommand{\textlineskip}{\baselineskip=13pt}
\newcommand{\smalllineskip}{\baselineskip=10pt}

\def\eightcirc{
\begin{picture}(0,0)
\put(4.4,1.8){\circle{6.5}}
\end{picture}}
\def\eightcopyright{\eightcirc\kern2.7pt\hbox{\eightrm c}}

\def\abstracts#1#2#3{{
    \centering{\begin{minipage}{4.5in}\baselineskip=10pt\footnotesize
    \parindent=0pt #1\par
    \parindent=15pt #2\par
    \parindent=15pt #3
    \end{minipage}}\par}}

\renewenvironment{thebibliography}[1]
    {\frenchspacing
     \ninerm\baselineskip=11pt
     \begin{list}{\arabic{enumi}.}
        {\usecounter{enumi}\setlength{\parsep}{0pt}
     \setlength{\leftmargin 12.7pt}{\rightmargin 0pt} 
    \setlength{\leftmargin 17pt}{\rightmargin 0pt}   
    \setlength{\leftmargin 22pt}{\rightmargin 0pt}   
         \setlength{\itemsep}{0pt} \settowidth
    {\labelwidth}{#1.}\sloppy}}{\end{list}}

\newcounter{itemlistc}
\newcounter{romanlistc}
\newcounter{alphlistc}
\newcounter{arabiclistc}

\newcommand{\fcaption}[1]{
        \refstepcounter{figure}
        \setbox\@tempboxa = \hbox{\footnotesize Fig.~\thefigure. #1}
        \ifdim \wd\@tempboxa > 5in
           {\begin{center}
        \parbox{5in}{\footnotesize\smalllineskip Fig.~\thefigure. #1}
            \end{center}}
        \else
             {\begin{center}
             {\footnotesize Fig.~\thefigure. #1}
              \end{center}}
        \fi}

\def\@citex[#1]#2{\if@filesw\immediate\write\@auxout
    {\string\citation{#2}}\fi
\def\@citea{}\@cite{\@for\@citeb:=#2\do
    {\@citea\def\@citea{,}\@ifundefined
    {b@\@citeb}{{\bf ?}\@warning
    {Citation `\@citeb' on page \thepage \space undefined}}
    {\csname b@\@citeb\endcsname}}}{#1}}

\newif\if@cghi
\def\citelow{\@cghifalse\@ifnextchar [{\@tempswatrue
    \@citex}{\@tempswafalse\@citex[]}}

\def\@refcitex[#1]#2{\if@filesw\immediate\write\@auxout
    {\string\citation{#2}}\fi
\def\@citea{}\@refcite{\@for\@citeb:=#2\do
    {\@citea\def\@citea{, }\@ifundefined
    {b@\@citeb}{{\bf ?}\@warning
    {Citation `\@citeb' on page \thepage \space undefined}}
    \hbox{\csname b@\@citeb\endcsname}}}{#1}}

\def\@refcite#1#2{{#1\if@tempswa\typeout
        {IJCGA warning: optional citation argument
    ignored: `#2'} \fi}}

\def\refcite{\@ifnextchar[{\@tempswatrue
    \@refcitex}{\@tempswafalse\@refcitex[]}}

\def\pmb#1{\setbox0=\hbox{#1}
    \kern-.025em\copy0\kern-\wd0
    \kern.05em\copy0\kern-\wd0
    \kern-.025em\raise.0433em\box0}

\def\fnt#1#2{\footnotetext{\kern-.3em
    {$^{\mbox{\scriptsize #1}}$}{#2}}}

\def\runninghead#1#2{\pagestyle{myheadings}
\markboth{{\protect\footnotesize\it{\quad #1}}\hfill}
{\hfill{\protect\footnotesize\it{#2\quad}}}}
\font\tenrm=cmr10
\font\tenit=cmti10
\font\tenbf=cmbx10
\font\bfit=cmbxti10 at 10pt
\font\ninerm=cmr9

\font\eightrm=cmr8

\def\qed{\hbox{${\vcenter{\vbox{            
   \hrule height 0.4pt\hbox{\vrule width 0.4pt height 6pt
   \kern5pt\vrule width 0.4pt}\hrule height 0.4pt}}}$}}

\begin{document}

\newpage

\runninghead{H.C. Rosu, O. Cornejo-P\'erez} {Factorization}

\normalsize\textlineskip
\thispagestyle{empty}
\setcounter{page}{1}


\vspace*{0.88truein}

\bigskip
\centerline{\bf FACTORIZATION OF DAMPED WAVE EQUATIONS WITH CUBIC NONLINEARITY}

\vspace*{0.035truein}
\vspace*{0.37truein}
\vspace*{10pt} \centerline{\footnotesize H. C.
ROSU\footnote{E-mail: hcr@ipicyt.edu.mx} , O. CORNEJO-P\'EREZ }
\vspace*{0.015truein}
\centerline{\footnotesize  Potosinian Institute of Science and Technology,}
\centerline{\footnotesize
Apdo. Postal 3-74 Tangamanga, 78231 San Luis Potos\'{\i}, Mexico }

\vspace*{0.225truein}

\centerline{\footnotesize Dated: Feb 8th, 2004} 

\vspace*{0.21truein} \abstracts{ {\bf Abstract.} The recent factorization scheme that we introduced for nonlinear polynomial ODEs in math-ph/0401040
is applied to the interesting case of damped 
wave equations with cubic nonlinearities. Traveling kink solutions are possible in the plane defined by the kink velocity versus the damping coefficient only 
along hyperbolas that are plotted herein.
 }{}{}


\textlineskip                  
\vspace*{12pt}                 

\vspace*{1pt}\textlineskip  
\vspace*{-0.5pt}
\noindent

PACS number(s):  
11.30.Pb

\noindent





\bigskip

\noindent
In the cubic nonlinear wave equation with damping \cite{ce86,l85}
\begin{equation}
\frac{d^2f}{dx_{0}^2}-\frac{d^2f}{dx_{1}^2}+\lambda_{0}\frac{df}{dx_{0}}-f+f^3=0\label{cubicdamp1}
\end{equation}
we set $x_{0}\rightarrow t$, $x_{1}\rightarrow x$ and 
$\tau=x-\alpha t$ to get the following equation 
\begin{equation}
(1-\alpha^2)\frac{d^2f}{d\tau^2}+\alpha
\lambda_{0}\frac{df}{d\tau}+f-f^3=0\label{cubicdamp2}~.
\end{equation}
Rewriting as
\begin{equation}
f^{\prime\prime}+\beta f^{\prime}+g(f)=0\label{cubicdamp3}~,
\end{equation}
where 
\begin{equation}\label{coef}
\beta=\frac{\alpha \lambda_{0}}{1-\alpha^2}, \qquad {\rm and} \qquad g(f)=f(1-f^2)~,
\end{equation}
we obtain the standard form of the equation allowing the application of our recent factorization scheme \cite{rcp04}.
In general, the factorization methods lead to traveling kink solutions, which are important in many applications. 
When first derivatives are present in a second-order differential equation with polynomial nonlinearity,  the kink solutions are obtained only for special 
values of the coefficient in front of the first derivative. For other values, more complicated solutions occur.
For example, in the case of Eq.~(\ref{cubicdamp1}), the general solutions are elliptic functions \cite{ce86,l85}. 

To apply the scheme in math-ph/0401040, we define $f_{1}$ and $\phi_{2}$ therein as the following functions
\begin{eqnarray}
f_{1}=\mp\frac{1}{\sqrt{2}}(1+f), \quad
\phi_{2}=\mp\sqrt{2}(1-f)\label{cubicdamp4}~.
\end{eqnarray}
Then, the factorization leading to the hyperbolic tangent kinks is possible for $\beta _{\rm factor}=\pm\frac{3}{\sqrt{2}}$.
The two different values obtained for $\beta$ yield the following
equation
\begin{equation}
\alpha^2\pm\frac{\sqrt{2}}{3}\lambda_{0}\alpha-1=0
\label{cubicdamp5}
\end{equation}
and solutions of Eq. (\ref{cubicdamp5}) as a function of
$\lambda_{0}$ are given by
\begin{equation}
\alpha _{1,2}=\frac{1}{2}\left(
-\frac{\sqrt{2}}{3}\lambda_{0}\pm\sqrt{\frac{2}{9}\lambda_{0}^2+4}\right)
\label{cubicdamp6}
\end{equation}
that is plotted in Fig.~(1), and
\begin{equation}
\alpha _{3,4}=\frac{1}{2}\left(
\frac{\sqrt{2}}{3}\lambda_{0}\pm\sqrt{\frac{2}{9}\lambda_{0}^2+4}\right)
\label{cubicdamp7}
\end{equation}
plotted in Fig.~(2).

\medskip

\vskip 1ex
\centerline{
\epsfxsize=270pt
\epsfbox{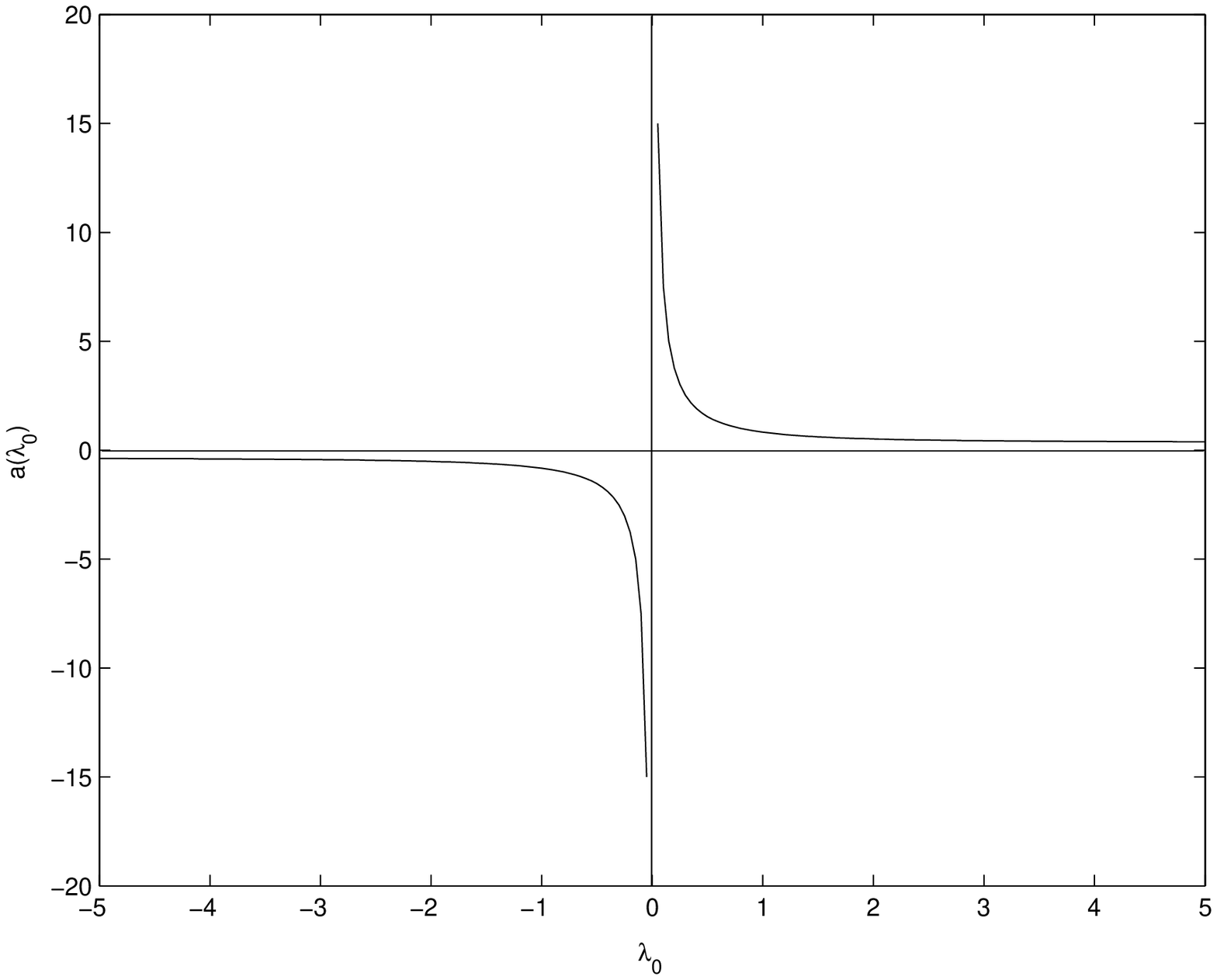}}
\vskip 2ex
\begin{center}
{\small{Fig. 1}$\,$
The curves $\alpha _{1,2}$ in the plane ($\alpha$,$\lambda _0$) along which the factorization is possible.}
\end{center}

\medskip

\vskip 1ex
\centerline{
\epsfxsize=270pt
\epsfbox{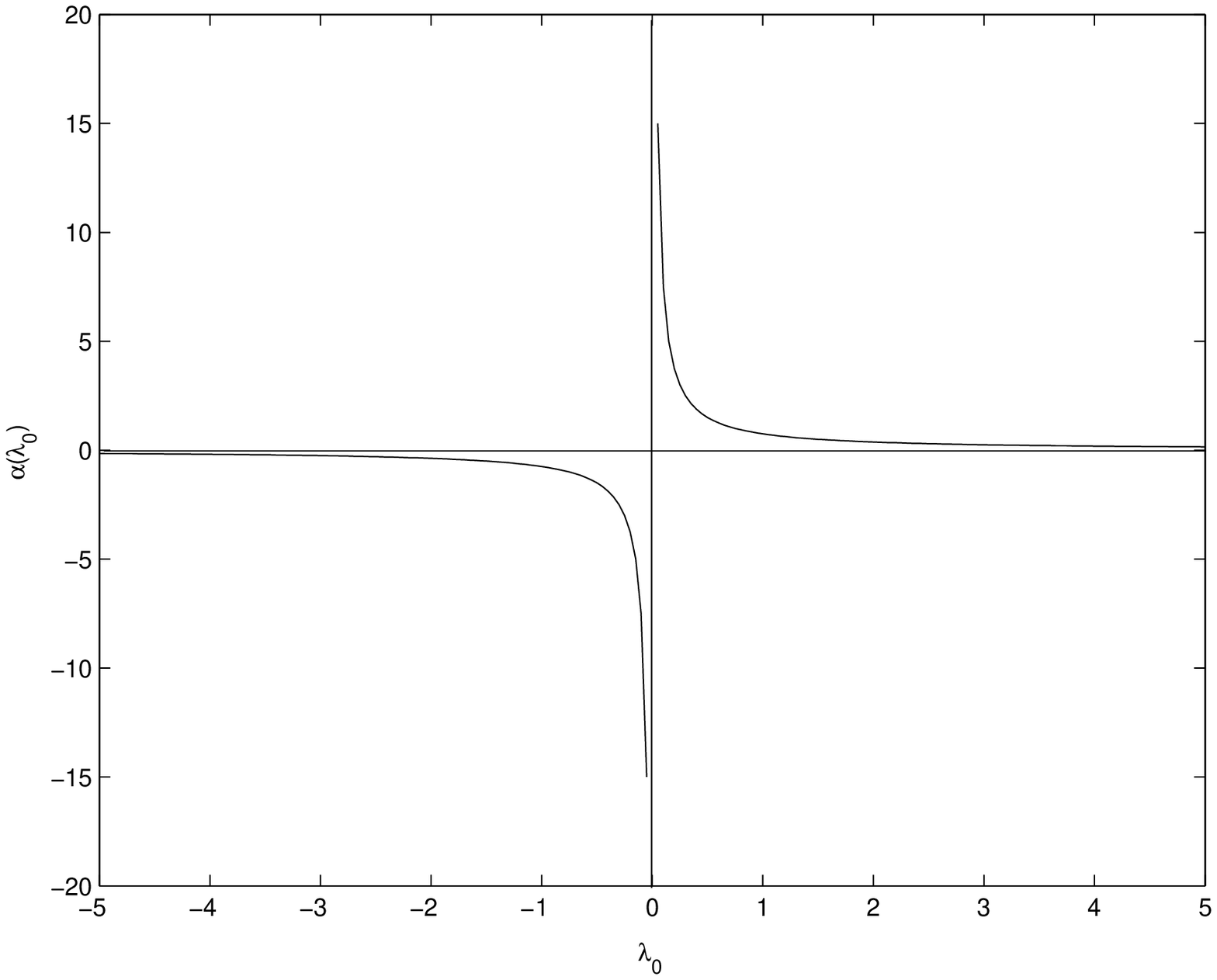}}
\vskip 2ex
\begin{center}
{\small{Fig. 2}$\,$
The curves $\alpha _{3,4}$ in the plane ($\alpha$,$\lambda _0$) along which the factorization is possible.}
\end{center}

\end{document}